\documentclass{PoS}
\usepackage{graphicx}

\newcommand{\nl}{\nonumber\\}

\newcommand{\be}{\begin{equation}}
\newcommand{\ee}{\end{equation}}
\newcommand{\bea}{\begin{eqnarray}}
\newcommand{\eea}{\end{eqnarray}}

\newcommand{\bbar}{B_q^0 - \overline {B^0_q}}

\newcommand{\msbar}{{\overline{\rm MS}}}

\title{Heavy flavor physics from lattice QCD}

\ShortTitle{Heavy Flavor Physics}

\author{Tetsuya Onogi\thanks{YITP-06-54}\\
        Yukawa Institute for Theoretical Physics, 
        Kyoto University, Kyoto 606-8502, Japan\\
        E-mail: \email{onogi@yukawa.kyoto-u.ac.jp}}

\abstract{I review the recent status of heavy flavor physics results
        from lattice QCD. In particular, I focus on the heavy-light
        decay constants, the bag parameters, the form factors, and the
        bottom quark mass. New progresses in theoretical  methods
        are also reviewed.}  

\FullConference{XXIV International Symposium on Lattice Field Theory\\
		 July 23-28 2006\\
		 Tucson Arizona, US}

\begin{document}

\section{Introduction}

There has been significant experimental progress owing to the remarkable
success in B factories. Recently there appeared measurements of the
mass difference $\Delta m_{B_s}$ from CDF~\cite{Abulencia:2006ze}
, Belle measurements of the pure leptonic decay $B\rightarrow \tau
\nu$ ~\cite{Ikado:2006un} , and the FCNC $B\rightarrow \rho , \omega
\gamma$ . Also, $\sin(2\phi_1)=sim(2\beta)$was measured with improved
precisions. The semileptonic inclusive and exclusive decays
$b\rightarrow c,u$ were also measured with much higher accuracies. We
can therefore overconstrain the CKM matrix elements with the present
experimental data. This will be a good test for QCD calculation , the
standard model, and the physics beyond the standard model.   

The CP asymmetry $A_{CP}(B\rightarrow J/\psi K)$ , the  mass difference $\Delta
m_{B_s,d}$, the branching fraction of the pure leptonic decay ${\cal
B}(B\rightarrow\tau^-\nu_{\tau})$,  and differential decay rates for
various semileptonic B decays can be written as
\begin{eqnarray}
A_{CP}(B\rightarrow J/\psi K)
&\propto& \sin(2\phi_1) = \sin(2\beta)\nonumber\\
\Delta m_{B_s} 
&=& \mbox{(known factors)} \  m_{B_s} f_{B_s}^2 
    \hat{B}_{B_s}|V_{ts}V_{tb}|,\nonumber\\
   \frac{\Delta m_{B_s}}{\Delta m_{B_d}}
&=& \frac{|V_{ts}|^2}{|V_{td}|^2} \frac{m_{B_s}}{m_{B_s}} 
    \frac{f_{B_s}^2 B_{B_s}}{f_{B_d}^2 B_{B_s}},\nonumber\\
{\cal B}(B\rightarrow\tau^-\nu_{\tau})
&=& \frac{G_F^2 m_B m_{\tau}^2}{8\pi}
    (1-\frac{m_{\tau}^2}{m_B^2})^2 f_B^2 |V_{ub}|^2 \tau_B, \nonumber\\
\frac{d\Gamma(B\rightarrow D^{(*)} l\nu)}{dw}
&=& \mbox{(known factors)} \ |V_{cb}|^2
\left\{ 
\begin{array}{@{\,}ll}
(w^2-1)^{1/2}  F_*^2(w) &\mbox{ For $B\rightarrow D^*$}\nonumber\\
(w^2-1)^{2/2}  F^2(w)   &\mbox{ For $B\rightarrow D$}
\end{array}
\right.,\nonumber\\
\frac{d^3\Gamma(B\rightarrow X_c l\nu)}{dE_l dq^2 d m_X^2}
&=& \mbox{(known factors)} \ |V_{cb}|^2
m_b^5 [1 +\frac{(\mbox{function of $\lambda_1$, $\lambda_2$)}}{m_b^2} +
\cdots ],
\nonumber\\
\frac{d\Gamma(B\rightarrow\pi l \nu)}{dq^2}
&=& \frac{G_F^2}{24\pi^3}
|(v \cdot k_{\pi})^2-m_{\pi}^2|^{3/2} |V_{ub}|^2 |f^+(q^2)|^2,\nonumber\\
\frac{d^3\Gamma(B\rightarrow X_u l\nu)}{dE_l dq^2 d m_X^2}
&=& \mbox{(known factors)} \ |V_{ub}|^2
m_b^5 [1 +\frac{\mbox{(function of $\lambda_1$, $\lambda_2$)}}{m_b^2} +
\cdots ].
\nonumber
\end{eqnarray}
The unitarity of the CKM matrix implies that $|V_{tb}|=1+O(\lambda^4)$
and $|V_{ts}|=|V_{cb}|( 1 + O(\lambda^2)) $, $|V_{ub}|=|V_{cb}|\lambda
(\rho-i\eta)$ and $|V_{td}|=|V_{cb}|\lambda
(1-\rho-i\eta+O(\lambda^2))$ using the Wolfenstein
parameterization. Thus, there are 11 independent experimental data for 3
unknown CKM parameters $|V_{cb}|$, $\rho$ and $\eta$ in Wolfenstein
parameterization. First,  $\sin(2\phi_1)=\sin(2\beta)$, which is a
function of $\rho$ and $\eta$ can be determined purely from
experiment. In order to determine the CKM parameters from other
channels, we need to know the hadronic parameters: the decay constants
$f_{B_s}$, $f_{B_d}$, the Bag parameters $B_{B_d}$, $B_{B_s}$,  and
the semileptonic form factors $F$, $F_*$, and $f^+$. HQET parameters
$m_b$, $\lambda_1$, $\lambda_2$ are also needed but they can be
determined from experiment alone using the moments in inclusive
semileptonic decays or rare decays.  

In fact,  we already have strong constraints from  $\Delta
m_{B_s}/\Delta m_{B_d}=17.31^{+0.33}_{-0.18}(stat)\pm0.07ps^{-1}$ ,
$\sin(2\phi_1)=\sin(2\beta)=0.69\pm0.03$, and
$|V_{cb}|=[4.45\pm0.045]\times 10^{-3}$ with inclusive  $B\rightarrow
X_c l\nu$ decay. 

\vspace{0.5cm}
\begin{figure}[h]
\begin{center}
\includegraphics[width=8cm]{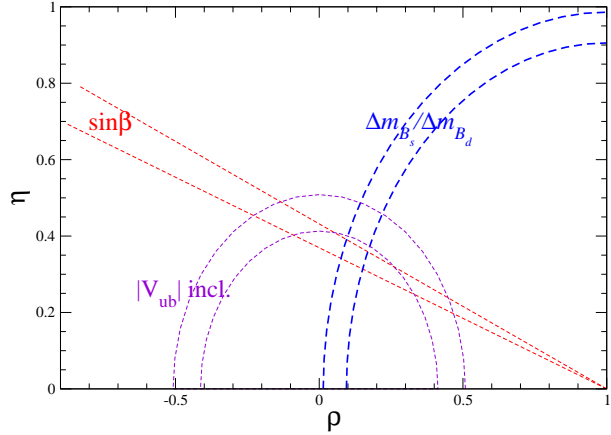}
\caption{Constraints on the unitarity triangle. 
The bands show bounds at 1 $\sigma$ level. }
\end{center}
\label{fig:CKM}
\end{figure}
As can be seen from Fig.~\ref{fig:CKM}, the results are consistent with
unitarity with 2 $\sigma$ level. However, it is also true that there
is still large room for new physics. Since the error is dominated by
theory except for $\sin(2\phi_1)=\sin(2\beta)$, it is crucial to
reduce the theoretical errors in the lattice determination of weak
matrix elements for heavy flavor physics for more stringent tests of
the standard model and the physics beyond.

\section{Heavy quark formalisms for heavy-light systems}

\subsection{Lattice NRQCD}
Lattice NRQCD action is a discretized version of nonrelativistic 
effective action which is applicable for the heavy quark mass 
whose spatial momenta are smaller than the mass. 
The expansion parameter is the velocity of the heavy quark for quarkonia and 
$\Lambda/m$ for the heavy-light system where $\Lambda$ is the 
typical momenta for all the light degrees of freedom. Since it 
is a nonrenormalizable theory, one cannot take the continuum 
limit. Also the action and operator can be matched to full QCD 
only by perturbation theory. In order to control the discretization 
errors the action is often highly improved at the tree level. 
The dominant source of errors are perturbative errors.

\subsection{Relativistic heavy quark (RHQ) formalism}
The Femilab action~\cite{El-Khadra:1996mp}
, AKT action~\cite{Aoki:2001ra}
, and the relativistic heavy quark(called RHQ) action ~\cite{Lin} 
by RBC collaboration are the formalisms for heavy quark using 
improved Wilson fermion with suitably chosen improvement coefficients. 
The three formulations are essentially the same in the sense that they 
are Synamzik effective action applicable to quarks with small spatial
momenta $|a\vec{p}|\ll1$ where the coefficients are mass dependent.
These actions smoothly interpolate the static quark and light quark. 
Therefore one can in principle take the continuum limit without 
encountering the breakdown of the theory. However, since the discretization
and perturbative error of the physical observable depend on $am$, 
 how the B meson physical observable approach to the continuum limit
is nontrivial.

The discretization and perturbative errors are expected to be small by
order estimation. Partial non perturbative (wavefunction)
renormilzation using $Z_V^{nonpert}$  is useful.
For higher accuracy both the discretization and perturbative error
should be reduced. In order to reduce the perturbative error
either two-loop calculation which is possible only by automated 
procedure~\cite{Nobes:2003nc} or nonperturbative renormalization.
To reduce the discretization error further improvement by adding 
more terms is necessary. This is perused by FNAL, CP-PACS and RBC
collaborations.

Lin and Christ~~\cite{Lin:2005ze},\cite{Lin} determined the
coefficients of the RHQ action nonperturbatively in quenched QCD.  
\begin{eqnarray}
S &=& \sum_n\bar{\psi}_n 
     [ m_0 +\gamma_0 +\zeta \vec{\gamma}\cdot\vec{D} 
     - \frac{1}{2} aD_0^2 -\frac{\zeta}{2}\vec{D}^2 
   -\sum_i \frac{i}{2}c_E a\sigma_{0i}F_{0i} 
     -\sum_{i,j} \frac{i}{2}c_B a\sigma_{i}F_{ij} ]\psi_n
\end{eqnarray}
They show that one can set $c_E=c_B=c_P$ by shifting $c_E$ and $c_B$
by field transformations 
\begin{eqnarray}
\psi \rightarrow (1-a^2[\gamma^i,\gamma^0][D^i,D^0]\xi_E)\psi,\\
\psi \rightarrow (1-a^2[\gamma^i,\gamma^u][D^i,D^j]\xi_B)\psi,
\end{eqnarray}
so that only three parameters$m_0$, $\zeta$ , $c_P$ should be tuned. 

In order to determined the parameters nonperturbatively, they carry
out the step scaling in three steps.
\begin{table}[h]
\begin{center}
\begin{tabular}{ll|ll|ll}
\hline
     &      &\multicolumn{2}{c}{finer lattice} &
     \multicolumn{2}{c}{coarser lattice}\\ 
\hline
     &L(fm) & size   & $a^{-1}_{finer}$ & sinze &
     $a^{-1}_{coarser}$(GeV)  \\ 
\hline
Step 1 &0.9   & $24^3\times 48$ & 5.4 GeV & $16^3\times 32$ & 3.6  \\
Step 2 &1.3   & $24^3\times 48$ & 3.6 GeV & $16^3\times 32$ & 2.4  \\
Step 3 &2.0   & $24^3\times 48$ & 2.4 GeV & $16^3\times 32$ & 1.6  \\
\hline
\end{tabular}
\end{center}
\caption{lattice setup for step scaling}
\end{table}
In step 1, one starts with a very fine lattice in small volume on
which $am \ll 1$ is satisfied so that one can describe the heavy quark
using Domain Wall fermion with controlled discretization error. 
One can then match the coefficients of the RHQ action on a coarser
lattice for the same volume using one shell quantities: (1) the spin
averaged 1S state mass for heavy-heavy and heavy-light system,
(2) hyperfine splitting for heavy-heavy and heavy-light system,
(3) the spin-orbit average and splitting for heavy-heavy system, and 
(4) the dispersion relation. 
In step 2, 3 and so on, they can repeat similar procedure to match 
RHQ on a lattice to  RHQ on an even coarser lattice. They demonstrate
that one can actually determine the parameters with reasonable
accuracy and obtain improvements in charmonium spectrum compared to
those with perturbatively determined parameters. This method is quite
similar to nonperturbative HQET by Alpha collaboration which will be
explained later. However, at the moment the step scaling function is
defined not in the continuum limit but a fixed lattice spacing assuming 
discretization error is under control. It will be important to have
theoretical understanding about how the systematic errors in the matching 
procedure can be controlled in this method.   

\subsection{Method with nonperturbative accuracy}

Rome II group~\cite{deDivitiis:2003iy},\cite{deDivitiis:2003wy}
proposed  a method to compute B physics observables 
with nonperturbative accuracy based on finite size scaling. 
Consider a physical observable $ {\cal O}(E_h,E_l)$ which depends on 
two largely separated energy scales $E_l$ and $E_h (E_l\ll E_h)$.
They assume that the finite size effects has a mild dependence on 
high energy scale $E_h$. Then Finite size effects can be obtained 
from the ratio $\sigma_{\cal O}$ of the observable in two different
volume $L$ and $2L$.
\begin{eqnarray}
\sigma_{\cal O}(E_l, E_h,L) =\frac{{\cal O}(E_l,E_h,2L)}{{\cal O}(E_l,E_h,L)}
\end{eqnarray}
When $E_l\ll E_h$ the finite size correction can be expanded as
\begin{eqnarray}
\sigma_{\cal O}(E_l, E_h,L) 
=\sigma_{\cal O}(E_l,L)
  + \frac{\alpha^{(1)}(E_l,L)}{E_h}
  + \frac{\alpha^{(2)}(E_l,L)}{E_h^2} + \cdots. 
\label{eq:1/Eh}
\end{eqnarray}
In the case of heavy-light meson almost at rest, the high energy
quantity $E_h$ is the heavy quark mass and the assumption that 
one can expand the physical observable ratio in $1/m$ is justified by HQET.
Using the step scaling function $\sigma_{\cal O}$ one can obtain the 
physical observable in infinitely large volume as 
\begin{eqnarray}
{\cal O}(E_l,E_h,L_{\infty})
= {\cal O}(E_l,E_h,L_0)
   \sigma_{\cal O}(E_l, E_h,L_0)
   \sigma_{\cal O}(E_l, E_h,2L_0)
   \cdots
\end{eqnarray}
When the volume is $L_0$ small one can carry out lattice
calculation with a cut off much larger than $E_h$  with reasonable
numerical cost so that one can compute the physical observable
directly at energy scale $E_h$ using the formalism of
nonperturbatively O(a)-improved Wilson fermion. But as the volume gets
larger through step scaling at some point $2^k L_0$ becomes too large
one cannot afford very small lattice spacing so that direct
computation becomes hopeless. However, one can always find a lower
energy scale $E_h^{(k)} < E_h $  where direct calculation is possible. 
In this case one can use Eq.~\ref{eq:1/Eh} to extrapolate  
$\sigma_{\cal O}(E_l,E_h^{(k)},2^k L_0) $ to $\sigma_{\cal O}(E_l,
E_h,2^k L_0) $. Since each step can be extrapolated in the continuum
with nonperturbatively O(a)-improved Wilson fermion, the only
systematic error in this procedure is the extrapolation in $1/E_h$. 
However, in order to take the continuum limit one has to know the 
parameter of constant physics so that one should know the master
formula $\lambda_{QCD}$ scale and renormalization invariant quark mass 
as functions of the bare gauge coupling $g^2_0$ and the bare quark mass
$m_0$. 
They find that in the case of the mass and the decay constant of the 
B meson mass one can practically control the extrapolation error at
the level of few percent accuracy. The advantage is that this method
is simple and promising. Probably, the Bag parameters, and form
factors at zero recoil also fall into this category. Form factors for
non zero recoil may be challenging. 

The Alpha collaboration proposes HQET with nonperturbative accuracy
~\cite{Heitger:2003nj} for high precision computation in B
physics. The action of HQET can be written with 1/m expansion as
follows  
\begin{eqnarray}
L = L_{stat} + \sum_{\nu}^n L^{(\nu)}, &\\
L_{stat} = \bar{\psi}_h [ D_0 + \delta m ] \psi_h, & L^{(\nu)} = \sum_i
\omega_i^{(\nu)} L_i^{(\nu)} 
\end{eqnarray}
$L_i^{(\nu)}$ are the $1/m^{\nu}$ correction terms
$\omega_i^{(\nu)}$ are their coefficients. 
\begin{eqnarray}
L_1^{(1)}= \bar{\psi}_h [ -\frac{1}{2} \sigma \cdot B ] \psi_h, 
&
L_2^{(1)}= \bar{\psi}_h [ -\frac{1}{2} D^2 ] \psi_h.
\end{eqnarray}
Since the static theory has a continuum limit and is a renormalizable
theory, if we expand the 1/m correction terms systematically to a
fixed order $n$ as operator insertions, one can renormalize all the
physical observable and take the continuum limit. In a very 
small volume where one can afford sufficiently fine lattice, the
renormalization parameter can be determined nonperturbatively by 
carrying out QCD simulation for heavy quark with $O(a)$-improved
Wilson  and imposing the matching condition for a set of physical
observables $\{\Phi_k(M,L_0)\}$  as,   
\begin{eqnarray}
\Phi_k^{HQET}(M,L_0) = \Phi_k^{QCD}(M,L_0)+O(\frac{1}{M^{n+1}}), 
 k=1,...,N_n 
\end{eqnarray}

After matching QCD to HQET in small volume with lattice size $L_0$, 
one can then define the step scaling function $F_k$ as
\begin{eqnarray}
\Phi_k^{HQET}(M,2L_0) = F_k(\Phi_k^{HQET}(M,L_0)) +O(\frac{1}{M^{n+1}}), 
 k=1,...,N_n 
\end{eqnarray}
By repeating this step,  one can determine the matching conditions 
for $\omega_i^{(\nu)}$ for large and coarse lattices where one wants 
to carry out lattice simulation.  During each step one can take the
continuum limit so that the only systematic error is the truncation
error in $1/M$. To control the truncation error $1/M \ll L_0 $ is
required which restricts the smallest possible $L_0$ as a function of
$M$.   This is in principle possible, but when one goes to higher order
mixing with lower dimension operators through power divergences may
give numerical difficulty, so that the calculation is technically more
demanding. 

In this conference  Guazzini et al.~\cite{Guazzini:2006bn}
reported their proposal for further improvements. They combine the
Rome II method and Alpha collaboration method. To be more precise,
they basically follow the Rome II method, but hey also compute step
scaling function $\sigma$ using nonperturbative HQET in the static
limit. When they estimate the heavy quark mass dependence of the
finite size correction, instead of extrapolating in 1/M,  they make
interpolation using the static result as an additional input.  

\section{Heavy-light decay constants}

\subsection{$f_{D_s}$, $f_{B_s}$ in quenched QCD}
The determination of the heavy-light decay constants with
nonperturbative accuracies is one of the most important progress.

Since the charm quark is of order 1 GeV, the decay constant $f_{D_s}$
in quenched approximation can be computed including nonperturbatively 
including the continuum limit with the present computer resources. 
Alpha collaboration~\cite{Juttner:2003ns}'s result for $a^{-1}=2-4$
GeV with $O(a)$-improved Wilson fermion is
\begin{eqnarray}
f_{D_s} = 252(9) \mbox{ MeV }.
\end{eqnarray}

The Rome II group~\cite{deDivitiis:2003wy}
 computed the heavy-light decay constants in quenched
QCD using $O(a)$-improved Wilson fermion by step scaling method. 
The observable is the nonpertubatively improved heavy-light axial
vector current in SF boundary for vanishing  boundary gauge field 
with periodic spatial boundary condition for fermions. 
They prepared three different size for $L^3 \times 2L$ volume 
with $L_0=0.4, L_1=0.8, L_2=1.6$ fm for step scaling. 
The lattice spacings and RGI heavy quark masses 
are  $a=0.012-0.033fm$, $m^{RGI}=1.6-7.0$ GeV for $L=L_0$,
$a=0.05-0.10fm$, $m^{RGI}=2.0-3.5$ for $L=L_1$ and  $a=0.10-0.20fm$,
$m^{RGI}=1.3-2.0$ for $L=L_2$ .  Defining the finite volume
corrections factors with the ratio of the decay constants for two
different volumes as  $\sigma(L_0) \equiv
\frac{f_{B_s}(2L_0)}{f_{B_s}(L_0)}$ and  $\sigma( L_1) \equiv
\frac{f_{B_s}(L_{\infty})}{f_{B_s}(2 L_0)}$ , the decay constant in
the infinite volume can be obtained as   
\begin{eqnarray}
f_{B_s}(L_{\infty}) = f_{B_s}(L_0)  \sigma( L_0) \sigma( L_1) .
\end{eqnarray}
The result is 
\begin{eqnarray}
f_{B_s}(L_0) = 475(2) \mbox{MeV}, & f_{D_s}(L_0) = 644(3) MeV\\
\sigma_{B_s}(L_1) = 0.417(3), &  \sigma_{D_s}(L_1) = 0.414(3)\\
\sigma_{B_s}(L_1) = 0.97(3), &   \sigma_{D_s}(L_1) = 0.90(2).
\end{eqnarray}
As it turned out, the heavy quark mass dependence of the step scaling
function are indeed small, which justified the
extrapolation. Combining these results
\begin{eqnarray}
f_{B_s} = 192(6)(4) \mbox{MeV}, & f_{D_s} = 240(5)(5) \mbox{MeV}.
\end{eqnarray}

Alpha collaboration~\cite{DellaMorte:2003mn}
compute static heavy-light decay constant with lattice HQET which is matched to QCD with nonperturbative accuracy by Schrodinger functional method. 
They computed the renormalization group invariant matrix element 
$\Phi^{stat}_{RGI}$ which can be related to the decay constant 
by a matching factor $C_{PS}$~\cite{Heitger:2003xg} as
$\Phi^{stat}_{RGI} = f_{PS}\sqrt{m_{PS}}/C_{PS}$ and obtain 
\begin{eqnarray}
\Phi^{stat}_{RGI}=1.74(13)
\end{eqnarray}
\vspace{0.5cm}
\begin{figure}[h]
\begin{center}
\includegraphics[width=8cm]{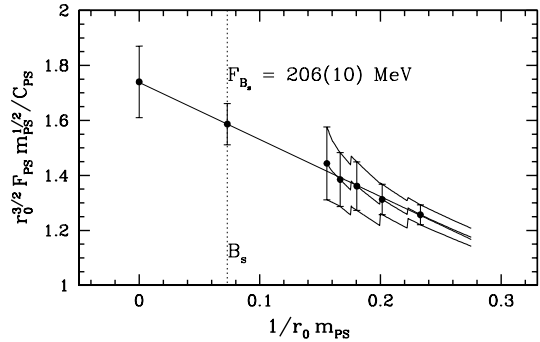}
\caption{Interpolation of static and relativistic results 
of heavy-light decay constant to obtain $f_{B_s}$. Figure taken from ~\cite{Rolf:2003mn}.}
\end{center}
\label{fig:fBs_alpha}
\end{figure}

Alpha collaboration~\cite{Rolf:2003mn} also computed the decay
constants for the charm quark mass regime, i.e. $m_Q=1.7-2.6$ GeV, at
four lattice spacings in the range $a=0.05-0.1$ fm  using
$O(a)$-improved Wilson fermion for both the heavy and the light quarks.  
They then interpolated the decay constants in the static limit and those 
for finite quark mass to obtain $f_{B_s}$. 
They found that both linear and quadratic interpolations lead to
\begin{eqnarray}
f_{B_s} = 206(10) \mbox{MeV}.
\end{eqnarray}
using $r_0=0.5$ fm for the scale input. 
as shown 
in Fig.2. 

Guazzini et al.~\cite{Guazzini:2006bn} reported a quenched study of
$f_{B_s}$  with nonperturbative accuracy. Their approach uses the
combination of the two methods. They compute the heavy-light decay
constants in finite volumes both for the  relativistic and static
heavy quarks to the step scaling  and ``interpolate'' the finite
volume correction $f_{B_s}$ to using both the relativistic and the
static.  They computed the 2 point functions for static and
relativistic heavy-light axial current
with Schrodinger boundary conditions, where the boundary gauge fields 
$C=C^{\prime}=0$ and periodic boundary condition in the spatial
directions $\theta=0$  for the light quark. The data for relativistic
heavy-light current is obtained by the reanalysis of those by Rome II
collaboration~\cite{deDivitiis:2003wy}.
They chose $f_{hl}\sqrt{m_{hl}}$  for the physical observable
rather than $f_{hl}$. Thus finite size corrections $\sigma_1$,
$\sigma_2$ are defined by the  ratio of $f_{hl}\sqrt{m_{hl}}$ for
different volumes as   
\begin{eqnarray}
\sigma_1
\equiv\frac{f_{hl}(2L_0)\sqrt{m_{hl}}(2L_0)}{f_{B_s}(L_0)\sqrt{m_{hl}}(L_0)}, 
&
\sigma_2 
\equiv \frac{f_{hl}(L_{\infty})\sqrt{m_{hl}}
(L_{\infty})}{f_{B_s}(2L_0)\sqrt{m_{hl}}(2L_0)}
\end{eqnarray}
the infinite volume can be obtained as   
\begin{eqnarray}
f_{B_s}(L_{\infty}) \sqrt{m_{B_s}(L_{\infty})} = f_{B_s}(L_0)  
\sqrt{m_{B_s}(L_0)}   \sigma_1 \sigma_2.
\end{eqnarray}

\vspace{0.5cm}
\begin{figure}[h]
\begin{center}
\vspace*{-15mm}
\begin{tabular}{ll}
\includegraphics[width=9.5cm]{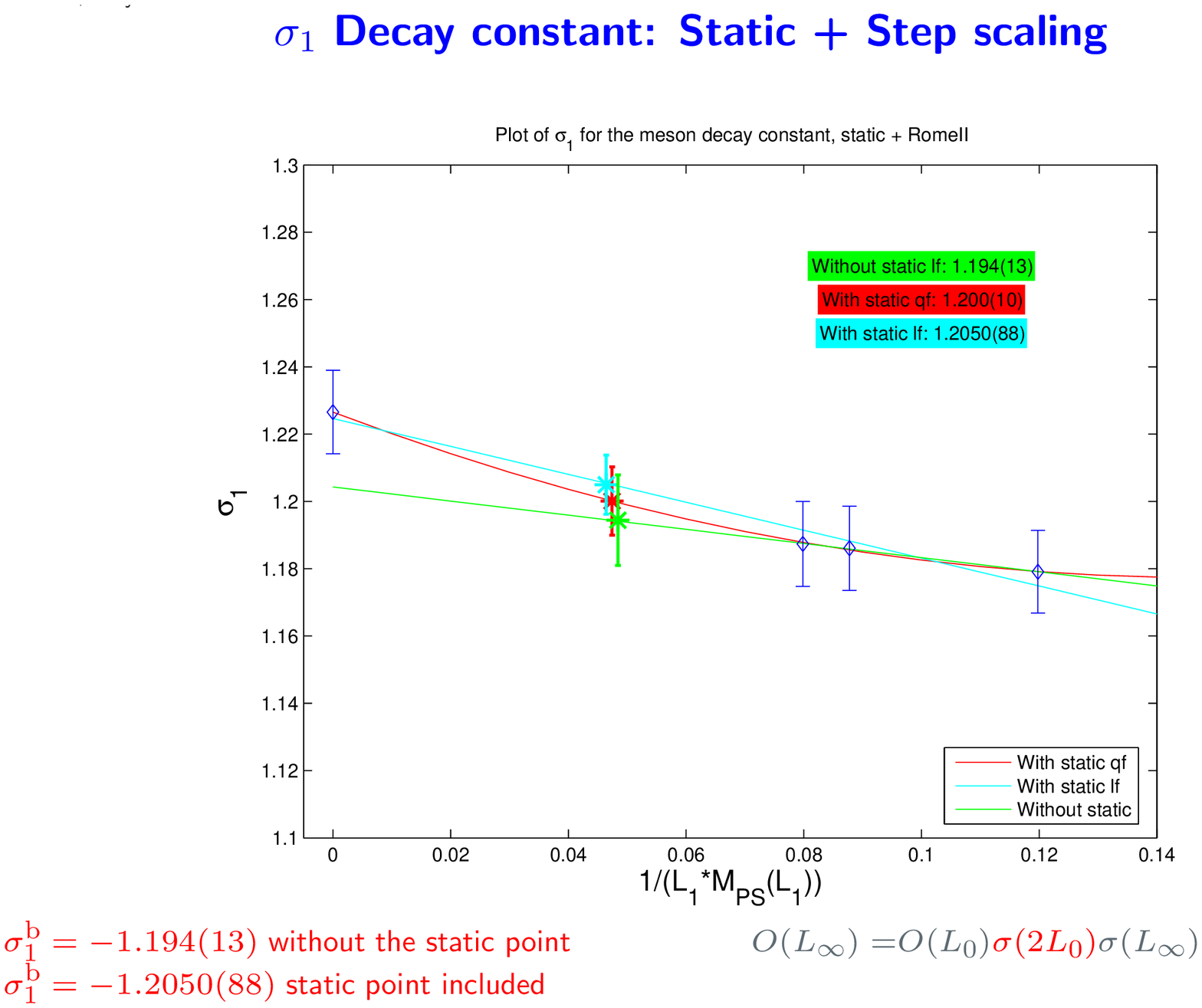} \\
\hspace*{-8mm}\vspace*{+5mm}
\includegraphics[width=11.5cm]{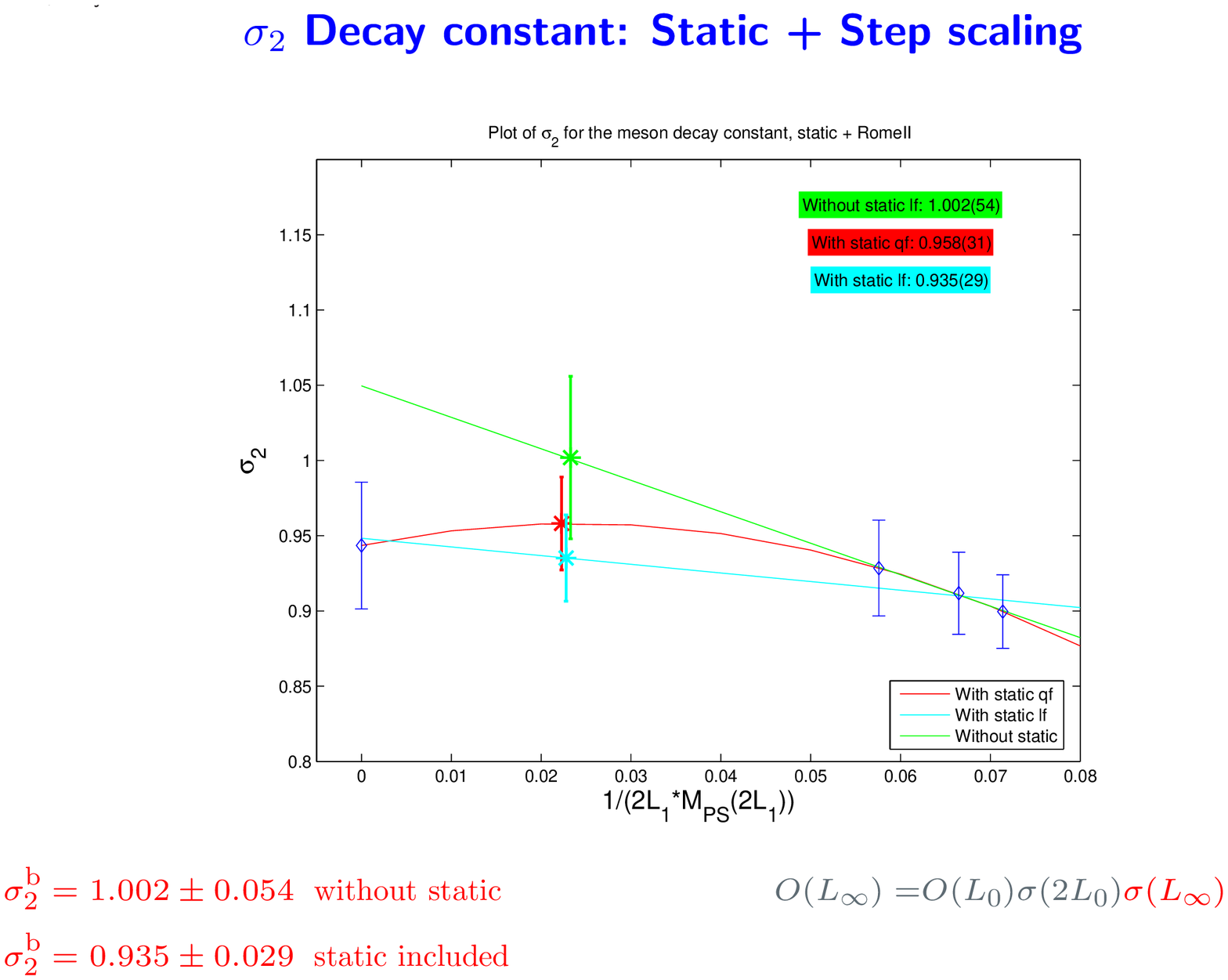}
\end{tabular}
\caption{1/M interpolation of the finite size corrections
$\sigma_1$(top) and $\sigma_2$(bottom) for
$f_{B_s}\sqrt{m_{B_s}}$. Figures from Guazzini's talk.}   
\end{center}
\label{fig: guazini_lat2006}
\end{figure}
As shown in 
Figs.3, 
the heavy quark mass dependences of the finite size corrections have
much better control with the help of static results. Their preliminary
quenched result is  
\begin{eqnarray}
f_{B_s} = 186 \pm 6  \mbox{ MeV } &  \mbox{ from Static + Rome II} \\
f_{B_s} = 195 \pm 11 \mbox{ MeV } &  \mbox{ from only Rome II}
 \end{eqnarray}
which are consistent with previous results by Rome II and by Alpha
collaborations.

There are also calculations of heavy-light decay constants with
Ginsparg-Wilson fermions. The RBC collaboration~\cite{Lin:2006vc} has
carried out a quenched study of D meson  using domain wall fermion
and DBW2 gauge action. The quark mass ranges from $m_q =\frac{1}{4}
m_s \sim \frac{5}{4} m_s$ and the lattice spacing is $a \sim 3$
GeV. Using the nonperturbative renormalization factor for the
light-light axial vector current~\cite{Aoki:2005ga} and giving the
mass correction as 
\begin{eqnarray}
Z_A^{hl}=Z_A^{ll,nonpert}
\frac{Z_{q,DWF}(am_{heavy})}{Z_{q,DWF}(am_{light})},
\end{eqnarray}
their result is  
\begin{eqnarray}
f_{D_s} &=& 254(4)(12) \mbox{ MeV},
\end{eqnarray}
where the errors are statistical, and systematic errors.
Chiu et al.~\cite{Chiu:2005gb}, \cite{Chiu:2005ue} also computed $f_D$
in quenched QCD using the optimal domain-wall fermion on a lattice 
with $a^{-1}=2.2$(GeV)  for 30 quark masses $am_q=0.03-0.80$ using
$f_{\pi}$ as scale input to find 
\begin{eqnarray}
f_{D_s} &=& 266(10)(18) \mbox{ MeV},
\end{eqnarray}
where the errors are statistical, and systematic errors.

\subsection{$f_{D_s}$, $f_{B_s}$ in unquenched QCD}

FNAL/MILC collaboration~\cite{ref:Simone} reported preliminary results
of $f_{B_s}$  
for $n_f=2+1$ flavor QCD with MILC configuration.
They use fermilab formalism for the heavy quark and improved staggered
for the light quark. The lattice spacings are $a=0.09 0.12, 0.15$ fm.
The renormalization factor $Z_A$ is taken to be
\begin{eqnarray}
Z^{Qq}_A = \rho_A^{Qq} \sqrt{Z_V^{QQ}Z_V^{qq}}, 
\end{eqnarray}
where $Z_V $'s are computed nonperturbatively and the remaining part
$\rho_A$ is computed by one-loop perturbation theory.
Their preliminary result is
\begin{eqnarray}
f_{B_s} = 253(7)(41) \mbox{ MeV}, & f_{B_s}/f_{D_s}  = 0.99(2)(6), 
\end{eqnarray}
where the errors are statistical error and systematic errors.

\vspace{0.5cm} 
\begin{figure}[h]
\begin{center}
\includegraphics[width=6cm]{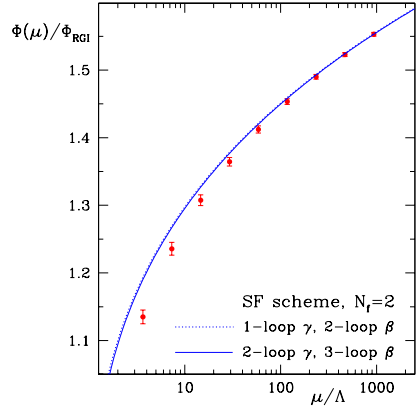}
\caption{Step scaling of heavy-light axial current. Figure provided
by J. Heitger.} 
\end{center}
\label{fig:ZA}
\end{figure}

Alpha collaboration~\cite{ref:Heitger}
computed the renormalization
factor  for the static heavy-light axial vector current $Z_A^{stat}$
for $n_f=2$ unquenched QCD. Their preliminary result is shown in
Fig. 4.
Their preliminary result is 
\begin{eqnarray}
\Phi(L_{max}) / \Phi_{RGI} = 1.14(1),
\end{eqnarray}
where $L_{max}$ is the physical lattice size in which one wants to
carry out the matrix element calculation.
Using this result, once the large volume $n_f=2$ unquenched
calculation $\beta=5.3$ 
for the regularization dependent renomalization factor 
$Z_A^{stat}(L_{max},g_0)$ and the lattice bare matrix element
$f_{B_s}^{stat} \sqrt{m_{B_s}})^{lat}(L_{max},g_0)$ is done,
 one can obtain the static heavy-light decay constant as
\begin{eqnarray}
f_{B_s}^{stat} \sqrt{m_{B_s}} = C_{PS}
 \frac{\Phi_{RGI}}{\Phi(L_{max})}
 Z_A^{stat}(L_{max}, g_0) 
(f_{B_s}^{stat} \sqrt{m_{B_s}})^{lat}(L_{max},g_0),
\end{eqnarray}
where $C_{PS}$ is perturbatively calculable conversion factor.
The large volume $n_f=2$ simulation is now in progress for $\beta=5.3$.

\subsection{Discussion on $f_{B_s}$, $f_{D_s}$ results}

\begin{figure}[h]
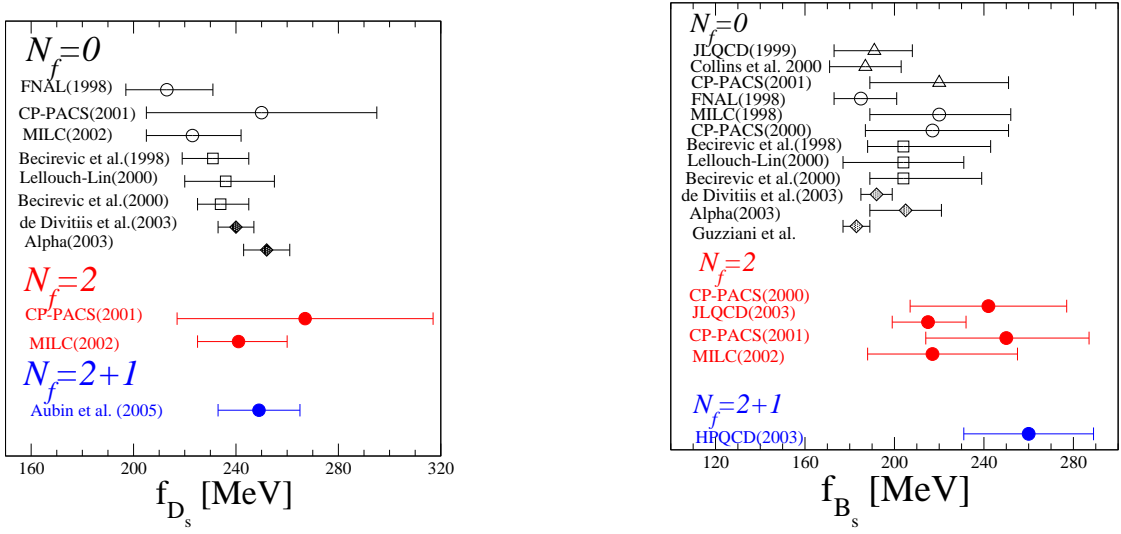

\vspace*{5mm}
\begin{center}
\begin{tabular}{lc r}
\includegraphics[width=6cm]{fDs.eps} & \hspace{2cm} &
\includegraphics[width=6cm]{fBs.eps} 
\end{tabular}
\caption{Decay constants $f_{D_s}$(left), $f_{B_s}$(right)} 
\end{center}
\label{fig:fDsBs}
\end{figure}

\vspace{0.6cm}
\begin{figure}[h]
\begin{center}
\includegraphics[width=8cm]{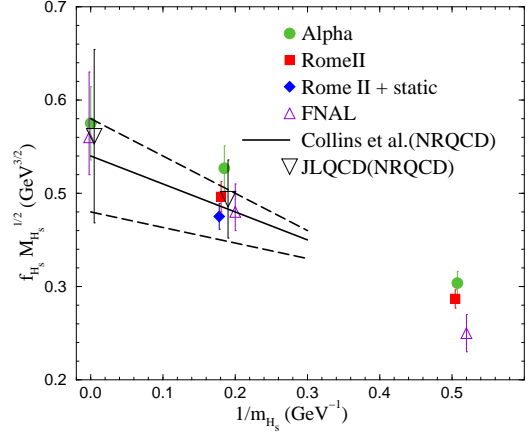}
\caption{Comparison of 1/m dependence of $f\sqrt{m}$}
\end{center}
\label{fig:Phi_compare}
\end{figure}

Fig. 5
show the summary of decay constants
$f_{B_s},f_{D_s}$ in quenched, $n_f=2$ unquenched,  and $n_f=2+1$
unquenched lattice QCD. It should be noted that the quenched
results are getting very precise owing to the recent developments 
with finite volume technique which allows us to compute
nonperturbatively renormalized heavy-light decay constants in the
continuum limit as discussed in the previous subsection.
We take the average of the results from  Rome II and Alpha
collaboration as the best result in quenched approximation, 
\begin{eqnarray}
f_{B_s}^{n_f=0} = 194(6) \mbox{  MeV}, &
f_{D_s}^{n_f=0} = 245(6) \mbox{  MeV}, & 
\left(\frac{f_{B_s}}{f_{D_s}}\right)^{n_f=0}=0.80(6)\nonumber,
\end{eqnarray}
In the unquenched case, the decay constants have larger errors
from perturbative matching. I would quote the average of HPQCD/MILC
and FNAL/MILC results for $f_B$ and FNAL/MILC results for $f_D$ as the
best value. However, since the best result come from the same
configuration, it would be worthwhile to study the heavy quark
mass dependence and $n_f$ dependence based on the collection of results
from various collaborations.   

\begin{table}[h]
\begin{center}
\begin{tabular}{lllll }
\hline\hline
$n_f$ & Group   &  heavy   & light & $\frac{f_{B_s}}{f_{D_s}}$ \\
\hline
0   & MILC~\cite{Bernard:1998xi} & Wilson   & Wilson   & 0.89(4) \\
    & FNAL~\cite{El-Khadra:1997hq}  & fermilab & clover   & 0.88(3) \\
    & Lellouch-Lin~\cite{Lellouch:2000tw} & clover   & clover   & 0.82(5)\\
    & Rome II~\cite{deDivitiis:2003wy}   & clover   & clover   & 0.80(4)\\
    & Alpha~\cite{Rolf:2003mn}       & clover   & clover   & 0.81(6)\\
    & RomeII+Alpha~\cite{Guazzini:2006bn} & clover   & clover   & 0.76(3)\\
\hline
2   & MILC~\cite{Bernard:2002pc} & Wilson   & Wilson   & 0.92(7) \\
\hline
2+1 & FNAL/MILC~\cite{ref:Simone}    & fermilab & Imp Stag & 0.99(2)(6) \\
\hline\hline
\end{tabular}
\caption{The decay constant ratio $f_{B_s}/f_{D_s}$. } 
\end{center}
\label{tab:fBsfDs}
\end{table}
Table 2  shows the ratio of $f_{B_s}/f_{D_s}$. 
Recent quenched calculations show smaller 
values of $\frac{f_{B_s}}{f_{D_s}}$. 
Fig. 6 shows
the comparison of $1/M$ dependence of $f_{B_s}\sqrt{f_{m_{B_s}}}$  in
quenched QCD near the static limit by Alpha, Rome II, FNAL, Collin's et
al. and JLQCD. It can be seen that the $1/M$ slope is consistently 
small independent of the action or collaborations.
Parameterizing 
\begin{eqnarray}
f_{B_s}\sqrt{f_{m_{B_s}}}= (f_{B_s}\sqrt{f_{m_{B_s}}})^{stat}
( 1 -\frac{c_1}{m_{B_s}} + \cdots ),
\end{eqnarray}
both the Alpha collaboration and Collins et al give the slope 
of $c_1\sim 0.5-0.6$ GeV.

MILC results for $n_f=2$  suggest that sea quark effects may increase
$\frac{f_{B_s}}{f_{D_s}}$ but not significantly due to the error. On
the other hand, FNAL/MILC preliminary $n_f=2+1$ result presented in
this conference suggests a significant increase in the
$\frac{f_{B_s}}{f_{D_s}}$.  However, one should bear in mind that the
the systematic error is slightly different for B and D in fermilab
formalism so that some consistency check is desired.

\begin{table}[h]
\begin{center}
\begin{tabular}{llllll}
\hline\hline
Group   &  heavy   & $a^{-1}$ input 
& $\frac{f_{B_s}^{n_f=2}}{f_{B_s}^{n_f=0}}$ 
& $\frac{f_{D_s}^{n_f=2}}{f_{D_s}^{n_f=0}}$ 
& $\frac{f_{B_s}^{n_f=2+1}}{f_{B_s}^{n_f=0}}$ \\
\hline
JLQCD ~\cite{Ishikawa:1999xu},~\cite{Aoki:2003xb}
        & NRQCD    & $m_{\rho}$ & 1.13(5) & - & - \\
CP-PACS~\cite{AliKhan:2001jg} 
& NRQCD    & $\sigma$   & 1.10(5) & - & -\\
HPQCD~\cite{Gray:2005ad}
   & NRQCD    & $r_0$      & -       & - & $\sim$ 1.15 \\
\hline
CP-PACS~\cite{AliKhan:2000eg}
& fermilab & $m_{\rho}$ & 1.14(5) & 1.07(5) & -\\
MILC~\cite{Bernard:1998xi},~\cite{Bernard:2002pc}    
& Wilson   & $f_{\pi}$  & 1.09(5) & 1.08(5) & -\\
\hline\hline
\end{tabular}
\caption{$n_f$ dependence of $f_{B_s}$, $f_{D_s}$. } 
\end{center}
\label{tab:nf_dep}
\end{table}

Table 3
is the collection of the $n_f$ dependence
of the heavy-light decay constants $f_{B_s}$, $f_{D_s}$ using the same
gauge and fermion action by the same group.  It is seen that if the
scale is set by the low energy inputs, turning on the sea quark
effects from $n_f=0$ to  $n_f=n+2$ to increases  $f_{B_s}$ by  10-15\%
, while the increase is not significant for $f_{D_s}$.  
It is quite natural to expect the size of the sea quark effects for
$f_{D_s}$ should be something between that for $f_{B_s}$ and $f_K$.
And with the low energy inputs $f_K$ receives almost no sea quark
effects by definition, the sea quark effects for $f_{B_s}$ should be
larger than for $f_{D_s}$, which explains the above observations. 

From Tables \ref{tab:nf_dep}, \ref{tab:fBsfDs}, 
we also estimate the ratio of decay constants as 
\begin{eqnarray}
\frac{f_{B_s}^{n_f=2}}{f_{B_s}^{n_f=0}}=1.12(5), &
\frac{f_{D_s}^{n_f=2}}{f_{D_s}^{n_f=0}}=1.08(5),\\ 
\frac{f_{B_s}^{n_f=2+1}}{f_{B_s}^{n_f=0}}=1.15(5), &
\frac{f_{D_s}^{n_f=2+1}}{f_{D_s}^{n_f=0}}=1.10(5).
\end{eqnarray}
This can give an educated guess for $n_f=2$ decay constants. 
However, there are several  uncertainties in this argument.
First, the up/down  sea quark mass for the unquenched configuration 
other than MILC may not small enough to fully reproduce the sea quark 
effects. Also when one uses low energy inputs the scale suffer from 
the chiral extrapolation uncertainty. Although Sommer scale $r_0$ is
relatively stable, but the phenomenological value of $r_0=0.5$ fm 
also suffer from uncertainty which is typically 10\%.  
Our educated guess for $n_f=2$ results are
\begin{eqnarray}
f_{B_s}^{n_f=2} = 217(12)(22) \mbox{  MeV}, &  
f_{D_s}^{n_f=2} = 265(14)(27) \mbox{  MeV}, \\
f_{B_s}^{n_f=2+1} = 223(17)(22) \mbox{  MeV}, &
f_{D_s}^{n_f=2} = 270(18)(27) \mbox{  MeV},   
\end{eqnarray}
where the second error is added to take account the scale
uncertainties of order 10\%. 
On the other hand the average based on the actual data of decay
constant in $n_f=2+1$ QCD by FNAL/MILC and FNAL/MILC collaborations are 
\begin{eqnarray}
f_{B_s}^{n_f=2+1} = 260(30) \mbox{  MeV},  &
f_{D_s}^{n_f=2+1} = 249(16) \mbox{  MeV},    
\end{eqnarray}
which is marginally consistent with our estimate within errors. 
Combining my educated guess and HPQCD/MILC, FNAL/MILC results 
my  'world average' would be 
\begin{eqnarray}
f_{B_s}^{n_f=2+1} = 240(30) \mbox{  MeV},  &
f_{D_s}^{n_f=2+1} = 260(20) \mbox{  MeV}. 
\end{eqnarray}

\subsection{chiral extrapolation}

In order to obtain $f_{B_d}$ and $f_{D_d}$ one has to take the chiral
extrapolation. This offers another important issue for precise
determination of the decay constant in addition to the problems
discussed for $f_{B_s}$ and $f_{D_d}$. The correct answer can only be
obtained with unquenched calculation.
The chiral perturbation theory tells us that 
the chiral logarithmic corrections to the SU(3) breaking ratio of the
decay constants is ~\cite{Grinstein:1992qt}
\begin{eqnarray}
{f_{B_s}\sqrt{m_{B_s}} \over f_{B_d}\sqrt{m_{B_d}}} 
=  1 + {1 + 3 {\hat g}^2\over 4 (4 \pi f)^2} \left(3
m_\pi^2\log {m_\pi^2\over \Lambda}
\right.
\left.- 2 m_K^2 \log {m_K^2\over
\Lambda}- m_\eta^2\log {m_\eta^2\over \Lambda} \right) 
+ \cdots .
\label{LOG1}
\end{eqnarray}

FNAL/MILC collaboration~\cite{ref:Simone} reported preliminary results
from $n_f=2+1$ heavy-light decay constants in the previous subsection. 
With the staggered quark the pseudoscalar mesons for each flavor
quantum number ($I$) has 16 tastes labeled by $\xi=P,A,T,V,I$. Their
masses are splitted as  
\begin{eqnarray}
m^2_I=(m_a+m_b)\mu + a^2 \Delta_{\xi}, & \xi=P,A,T,V,I 
\end{eqnarray}
The staggered chiral perturbation theory suggests that the quark mass
dependence of the heavy-light decay constant $\Phi_{H_q}\equiv f_{H_q}
\sqrt{m_{H_q}}$ is  
\begin{eqnarray}
\Phi_{H_q} = \Phi_{H} [ 1 + \frac{\delta f_{H_q}}{16\pi^2f^2} + \mbox{
analytic terms } ] , 
\end{eqnarray}
where the explicit form of $\delta f_{H_q}$, which is the analog of 
the chiral log in the  continuum chiral perturbation theory, can be
obtain from staggered chiral perturbation theory~\cite{Aubin:2005aq}.
Due to the taste symmetry breaking of $O(a^2)$ terms they have
many parameters for $a^2$ effects which have to be fitted from the
lattice spacing dependence of the lattice data. Some parameters can be
obtained from the pion system but other parameters have to be fitted
from the data of the heavy-light decay constants themselves.    
Their preliminary results are
\begin{eqnarray}
(f_{D_s}/f_{D_d})^{n_f=2+1} = 1.21(1)(4). 
& (f_{B_s}/f_{B_d})^{n_f=2+1} = 1.27(2)(6),
\end{eqnarray}
where the errors are statistical and systematic errors.

Gadiyak and Loktik~\cite{Gadiyak:2005ea} made a  $n_f=2$ unquenched
study of SU(3) breaking 
effect using domain wall fermion and DBW2 gauge action. The quark mass
ranges from $m_{\pi}=490,610,700$ MeV and the lattice spacing is $a
\sim 1.69(5)$ GeV.  They found that
\begin{eqnarray}
(f_{B_s}/f_{B_d})^{n_f=2} = 1.29(4)(4)(2).
\end{eqnarray}

\begin{table}[h]
\begin{center}
\begin{tabular}{llllll}
\hline\hline
Group & heavy & light & $n_f$ & $f_{B_s}/f_{B_d}$  & visible chiral log \\
\hline
CP-PACSS~\cite{AliKhan:2001jg}
        & NQCD     & clover   & 2  & 1.18(2)(2)                 & NO\\
CP-PACS ~\cite{AliKhan:2000eg} 
        & fermilab & clover   & 2  & 1.20(3)(3)($^{+4}_{-0}$)   & NO\\
MILC~\cite{Bernard:2002pc} 
       & fermilab & Wilson   & 2  & 1.16(1)(2)(2)($^{+4}_{-0}$)& NO\\
JLQCD~\cite{Aoki:2003xb}
        & NRQCD    & clover   & 2  & 1.13(3)($^{+13}_{-0}$)     & NO\\
Gadiyak and Loktik~\cite{Gadiyak:2005ea} 
        & static   & DW       & 2  & 1.29(4)(6)                 & NO\\
\hline
HPQCD/MILC~\cite{Gray:2005ad}
 & NRQCD    & Imp Stag &2+1 & 1.20(3)(1)             & YES\\
FNAL/MILCC~\cite{ref:Simone}
  & fermilab & Imp Stag &2+1 & 1.27(2)(6)             & YES\\
\hline\hline
\end{tabular}
\caption{SU(3) breaking ratio $f_{B_s}/f_{B_d}$}
\end{center}
\label{tab:fBSU(3)} 
\end{table}

\begin{table}[h]
\begin{center}
\begin{tabular}{llllll}
\hline\hline
Group & heavy & light & $n_f$ & $f_{D_s}/f_{D_d}$  & visible chiral log \\
\hline
CP-PACSS ~\cite{AliKhan:2000eg} 
 & fermilab & clover   & 2  & 1.18(4)(3)($^{+4}_{-0}$)  & NO\\
MILC~\cite{Bernard:2002pc}    
    & fermilab & Wilson   & 2  & 1.14(1)($^{+2}_{-3}$)
(2)($^{+5}_{-0}$)  & NO\\
\hline
FNAL/MILC~\cite{ref:Simone}  & fermilab  & Imp Stag &2+1 & 1.21(1)(4)       & YES\\
\hline\hline
\end{tabular}
\caption{SU(3) breaking ratio $f_{D_s}/f_{D_d}$}
\end{center}
\label{tab:fDSU(3)} 
\end{table}

\begin{figure}[h]
\begin{center}
\vspace{1.5cm}
\includegraphics[width=8cm]{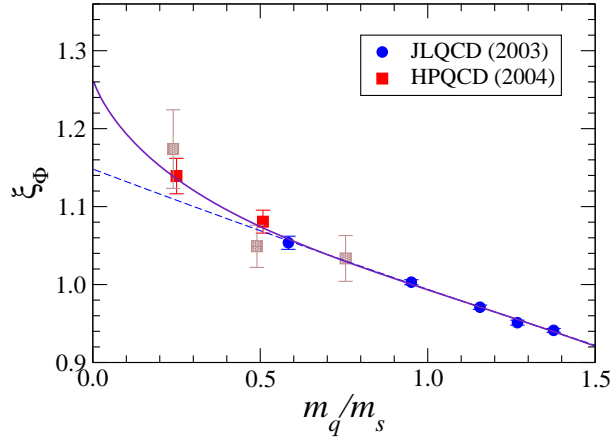}
\caption{light quark mass dependence of  $\Phi_{B_s}/\Phi_{B_q}$, 
where $\Phi=\sqrt{m_B} f_B$ .} 
\end{center}
\label{fig:xi}
\end{figure}

Tables 4,5  show the collections
of the unquenched results of $f_{D_s}/f_{D_d}$ and $f_{B_s}/f_{B_d}$.
Except for FNAL/MILC and HPQCD/MILC, they do not observe the chiral
log. This is natural because other results use much heavier light
quarks. 
Fig.7
show the comparison of the light quark mass
dependence of  $f_{B_s}\sqrt{m_{B_s}}/f_{B_d}\sqrt{m_{B_d}}$ from
JLQCD and HPQCD. They show consistent behavior for larger light quark
mass. It seems that the JLQCD result may be missing the possible onset
of chiral log which is found by HPQCD data. 
However, the results with MILC configuration are obtained
through the staggered chiral perturbation theory, which requires quite
complicated analysis with many parameters. Independent calculations
with other formalisms are needed.


\section{Bag parameters}
The bag parameters that parameterizes the $\bbar$ mixing amplitude are
defined by 
\begin{eqnarray}
 \label{Bparam}
\langle \bar B^0_q \vert \bar b^i \gamma_\mu (1- \gamma_{5} )  q^i \,
 \bar b^j  \gamma_\mu (1- \gamma_{5} ) q^j \vert   B^0_q \rangle  
&= {8\over 3} \, m_{B_q}^2  f_{B_q}^2 
 B_{B_q} &   \ (\mbox{ where } q=d,s),\\
\langle \bar B^0_s \vert \bar b^i (1- \gamma_{5} )  q^i \,
 \bar b^j  (1- \gamma_{5} ) s^j \vert   B^0_s \rangle  
&= -{5\over 3} \, m_{B_s}^2  f_{B_s}^2  
\frac{B_S}{R^2} & \  (\mbox{ where } R\equiv
\frac{(\bar{m_b}+\bar{m_s})}{m_{B_s}}), \\ 
\langle \bar B^0_s \vert \bar b^i (1- \gamma_{5} )  q^j \,
 \bar b^j  (1- \gamma_{5} ) q^i \vert   B^0_q \rangle  
&= {1\over 3} \, m_{B_q}^2  f_{B_q}^2
\frac{\tilde{B}_S}{R^2} &  .  
\end{eqnarray}

HPQCD~\cite{ref:Shigemitsu} 
computed the bag parameters for $B_s$ mixing calculation  with
improved $n_f=2+1$ dynamical staggered quark. The simulation
was carried out using NRQCD action for heavy  quark and AsqTad action
for light quark on a $20^3 \times 64$ lattice with $a^{-1}\sim 1.6$
GeV with the valence light quark mass at$m_s$ and the ud sea quark
mass at $0.25m_s$,  $0.5m_s$.
They computed the matrix elements for three types of $\Delta B=2$
four fermion operators which correspond to $f_{B_s}^2 B_{B_s}$,
$f_{B_s}^2 \frac{B_S}{R^2}$ and $f_{B_s}^2
\frac{\tilde{B}_S}{R^2}$. 

Defining $\Delta B=2$ four-fermion operators as
\begin{eqnarray}
OL & \equiv & [\bar{b}^i q^i]_{V-A} [\bar{b}^j q^j]_{V-A},\\
OS & \equiv & [\bar{b}^i q^i]_{S-P} [\bar{b}^j q^j]_{S-P},\\
Q3 & \equiv & [\bar{b}^i q^J]_{S-P} [\bar{b}^j q^II_{S-P},\\
OLj1 & \equiv & \frac{1}{2M}{ [\vec{\nabla}\bar{b}^i
\vec{\gamma}q^i]_{V-A} [\bar{b}^j q^j]_{V-A} + [\bar{b}^i q^i]_{V-A}
[\vec{\nabla}\bar{b}^j \vec{\gamma}q^j]_{V-A}},\\  
OSj1 & \equiv & \frac{1}{2M}{ [\vec{\nabla}\bar{b}^i
\vec{\gamma}q^i]_{S-P} [\bar{b}^j q^j]_{S-P} + [\bar{b}^i q^i]_{S-P}
[\vec{\nabla}\bar{b}^j \vec{\gamma}q^j]_{S-P}},\\ 
Q3j1 & \equiv & \frac{1}{2M}{ [\vec{\nabla}\bar{b}^i
 \vec{\gamma}q^j]_{S-P} [\bar{b}^j q^i]_{S-P} + [\bar{b}^i q^j]_{S-P}
[\vec{\nabla}\bar{b}^j \vec{\gamma}q^i]_{S-P}}, 
\end{eqnarray}
where $i,j$ are color indices. The former three are operators in the
leading order in $1/M$  and the latter three operators are $O(1/M)$ operators.
The lattice operator is matched to that in the continuum using
one-loop perturbation theory as 
\begin{eqnarray}
\frac{1}{2 m_{B_s}}\langle OL \rangle^{\bar{MS}}
&=& +  [1+\alpha\rho_{LL} ]\langle OL\rangle_{eff}
         +\alpha\rho_{LS} \langle OS\rangle_{eff}\\
& & +  [\langle OL\rangle_{eff}
       -\alpha(\zeta_{10}^{LL} \langle OL \rangle_{eff}
              +\zeta_{10}^{LS} \langle OS \rangle_{eff})],\\
\frac{1}{2 m_{B_s}}\langle OS \rangle^{\bar{MS}}
&=& +  [1+\alpha\rho_{SS} ]\langle OS\rangle_{eff}
         +\alpha\rho_{SL} \langle OL\rangle_{eff}\\
& & +  [\langle OSj1\rangle_{eff}
       -\alpha(\zeta_{10}^{SL} \langle OL \rangle_{eff}
              +\zeta_{10}^{SS} \langle OS \rangle_{eff})],\\
\frac{1}{2 m_{B_s}}\langle Q3\rangle^{\bar{MS}}
&=& +  [1+\alpha\rho_{33} \langle Q3\rangle_{eff}
         +\alpha\rho_{3L} \langle OL \rangle_{eff}\\
& & +  [\langle Q3j1\rangle_{eff}
       -\alpha(\zeta_{10}^{3L} \langle OL \rangle_{eff}
              +\zeta_{10}^{33} \langle Q3 \rangle_{eff})].
\end{eqnarray}
It should be noted that in this work dimension 7 operators are 
included for the first time in NRQCD. Previous work by Hiroshima
group~\cite{Hashimoto:2000eh} and JLQCD~\cite{Aoki:2002bh,Aoki:2003xb} 
include only dimension 6 operators.

They find that the sea quark mass is only a few percent and 
quote the result for $m_{sea}=0.25 m_s$ as their best value.
\begin{eqnarray}
f_{B_s}\sqrt{\hat{B}_{B_s}} = 0.281(20) \mbox{GeV}, &
f_{B_s}\sqrt{B_{B_s}}       = 0.227(16) \mbox{GeV},\\
f_{B_s}\sqrt{\frac{B_S}{R}} =  0.295(21) \mbox{GeV}, &
f_{B_s}\sqrt{\frac{\tilde{B}_S}{R}} = 0.305(21) \mbox{GeV}
\end{eqnarray}
The key points of HPQCD's result is that the direct calculation 
of $f_{B_s}^2 B_{B_s}$ gives better accuracy than computing 
$f_{B_s}$ and $B_{B_s}$ separately.  The bag parameter has a smaller
central value than that of JLQCD ($n_f=2$) after including $1/M$
correction (dime=7 operator), which is not included in JLQCD's
calculation. On the other hand, $f_{B_s}$ has a larger 
central value than JLQCD so that $f_{B_s}^2 B_{B_s}$ is consistent 
Table 6.

$f_{B_s}\sqrt{\hat{B}_{B_s}}$ is related to the 
mass difference in $B_s-\bar{B}_s$ mixing as 
\begin{eqnarray}
\Delta m_{B_s} 
&=& \frac{G_F^2}\eta_B m_{B_s} f_{B_s}^2 \hat{B}_{B_s}
    m_W^2 S_0(m_t^2/m_W^2)|V_{ts}V_{tb}|,
\end{eqnarray}
where $\eta$ is perturbatively calculable factor and
$S_0(m_t^2/m_W^2)$ is the Inami-Lim
function. CDF~\cite{Abulencia:2006ze} recently measured 
the mass difference as 
\begin{eqnarray}
\Delta m_{B_s}=18.3 (^{+4}_{-2}) \mbox{ps}^{-1}
\end{eqnarray}
Combining this with  recent $|V_{cb}|$ value and CKM unitarity
relation,  the above equation requires
$f_{B_s}\sqrt{\hat{B}_{B_s}}=0.245(20)$ (GeV).

Using heavy quark expansion  the width difference of $B_s-\bar{B}_s$
mixing can be obtained at NLO as 
\begin{eqnarray}
\left(\frac{\Delta \Gamma}{\Gamma}\right)_{B_s} 
& =& \frac{16\pi^2 B(B_s\rightarrow X
e\nu)}{g(m_c^2/m_b^2)\tilde{\eta}_{QCD}}   
\frac{f_{B_s}^2 m_{B_s}}{m_b^3} |V_{cs}|^2 \nonumber\\
& & \times \left(   G(z) \frac{8}{3}B_{B_s}(m_b) 
 + G_S(z) \frac{5}{3}\frac{B_{S_s}(m_b)}{R(m_b)^2}  
+ \sqrt{1-4mc^2/m_b^2} \delta_{1/m}  \right),
\end{eqnarray}
where $g(z)=1-8z+8z^3-z^4-12z^2 \ln z$ and $\tilde{\eta}_{QCD}$ is the
short distance QCD correction. $G(z)$ and  $G_S(z)$ are NLO QCD
corrections which appear in OPE. $\delta_{1/m}$ is the NLO
contribution in $1/m_b$ expansion. Using HPQCD results 
they predict 
\begin{eqnarray}
\left(\frac{\Delta \Gamma}{\Gamma}\right)_{B_s} 
= 0.16(3)(2),
\end{eqnarray}
where the second errors comes from the uncertainty in  the correction
term $\sqrt{1-4mc^2/m_b^2} \delta_{1/m} $.

\begin{table}[h]
\begin{center}
\begin{tabular}{llllll}
\hline
\hline
$n_f$ & group & heavy  & $B_{B_s}(m_b)$ & $B_{B_s}(m_b)$ & $B_{B_s}(m_b)$\\
\hline
0   &  Becirevic. et al.~\cite{Becirevic:2001xt} & HQET
 &  0.87(5)  &0.84(4) & 0.91(8) \\
0   & JLQCD~\cite{Aoki:2002bh}  & NRQCD & 0.84(5)      &0.85(5)   & -    \\
\hline
2   & JLQCD~\cite{Aoki:2003xb}  & NRQCD& 0.85(6)      & -      & - \\
\hline
2+1 & HPQCD~\cite{ref:Shigemitsu}  & NRQCD & 0.76(11)     &0.84(12)& 0.90(13) \\
\hline
\hline
\end{tabular}
\caption{The bag parameters and $B_{B_s}^\msbar (m_b)$ }
\end{center}
\label{tab:BBs}
\end{table}

\begin{table}[h]
\begin{center}
\begin{tabular}{llllll}
\hline
\hline
$n_f$ & group & heavy  & $B_{B_d}(m_b)$ \\
\hline
0   &  UKQCD   & HQET  &  0.87(5)  \\
0   &  Becirevic. et al.~\cite{Becirevic:2001xt} & HQET  &  0.87(6)  \\
0   & JLQCD~\cite{Aoki:2002bh}  & NRQCD & 0.84(6)      \\
2   & JLQCD~\cite{Aoki:2003xb}  & NRQCD & 0.84(6)       \\
\hline
\end{tabular}
\caption{The bag parameters and $B_{B_d}^\msbar (m_b)$ }
\end{center}
\label{tab:BBd}
\end{table}
Table 7 gives the summary of $B_{B_d}$ from various
collaborations. It should be noted that HPQCD's result with $1/m$
correction in the operator gives lower values. Further understanding 
of $1/m$ dependence is required. On the other hand, the light quark 
mass dependence seems small from the data.  In fact, chiral
perturbation theory~\cite{Grinstein:1992qt} suggests  that the light quark 
mass dependence is   
\begin{eqnarray}
\label{Bfac} {\hat B_{B_s} \over \hat
B_{B_d}} =1 + \frac{1 -3 {\hat g}^2}{(4 \pi f)^2} m_\pi^2\log m_\pi^2
+\cdots.  \;,  
\eea  
Since $g\sim 0.6$, The coefficient of the chiral log is very small,
which agrees with the lattice results.


\section{$B\rightarrow\pi l\nu$ form factors}
The matrix element 
$\langle\pi(k_{\pi})|\bar{q}\gamma_{\mu}b|B(p_B)\rangle$ 
for the heavy-to-light semi-leptonic decay $B\rightarrow\pi l\nu$ is
often parameterized as
\begin{equation}
  \label{eq:f+f0}
  \langle\pi(k_{\pi})|\bar{q}\gamma^{\mu}b|B(p_B)\rangle
  = f^+(q^2) 
  \left[ 
    (p_B+k_{\pi})^{\mu} 
    - \frac{m_B^2-m_{\pi}^2}{q^2} q^{\mu}
  \right]
  + f^0(q^2) \frac{m_B^2-m_{\pi}^2}{q^2} q^{\mu},
\end{equation}
with $p_B$ and $k_{\pi}$ the momenta  and $q = p_B-k_{\pi}$.
The differential decay rate of the semileptonic
$B^0\rightarrow\pi^- l^+\nu_l$ decay is 
\begin{equation}
  \label{eq:differential_decay_rate}
  \frac{1}{|V_{ub}|^2}
  \frac{d\Gamma}{dq^2} =
  \frac{G_F^2}{24\pi^3}
  |\vec{k}_{\pi}|^3
  |f^+(q^2)|^2.
\end{equation}
from which one can extract the CKM element $|V_{ub}|$.

HPQCD collaboration~\cite{Gulez:2006dt} has made a new study of 
$B\rightarrow\pi l\nu$ form factors using 2+1 flavor MILC
configuration with $a^{-1}=1.6, 2.3$ GeV. They used 
NQCD action for the heavy quark and improved staggered fermion 
for the light quark. The light quark mass ranges $m_q/m_s=0.125-0.5$
on the coarse lattice and $m_q/m_s=0.2-0.4$ on the fine lattice. 
The heavy-light vector current is renormalized with 1-loop matching 
through $O(\alpha/M)$ and $O(\alpha a)$. The chiral extrapolation is
carried out using staggered chiral perturbation theory. 
In order to make the analysis convenient they parameterize the matrix
element as
\begin{eqnarray}
\langle \pi(k_{\pi})| V^{\mu}| B(p_B) \rangle 
&= & \sqrt{2 m_B} [ v^{\mu} f_{\parallel} + k^{\mu}_{\perp} f_{\perp}],
\end{eqnarray}
with
\begin{eqnarray}
 v^{\mu} = \frac{p_B^{\mu}}{m_B}, & 
k^{\mu}_{\perp} = k_{\pi}^{\mu} -(k_{\pi}\cdot v) v^{\mu}.
\end{eqnarray}
In order to interpolate in $q^2$, they used several different pole
model fit functions. The first one is BK parameterization with three
parameters with $\tilde{q}^2 \equiv q^2/m_{B^*}$ , 
\begin{eqnarray}
f^+(q^2) = \frac{f(0)}{(1-\tilde{q}^2)(1-\alpha\tilde{q}^2)}, 
& f^0(q^2) = \frac{f(0)}{(1-\tilde{q}^2/\beta)}.
\end{eqnarray}
The second one is BZ parameterization with four parameters
\begin{eqnarray}
f^+(q^2) = \frac{f(0)}{(1-\tilde{q}^2)}
+ \frac{r \tilde{q}^2}{(1-\tilde{q}^2)(1-\alpha\tilde{q}^2)}, 
\end{eqnarray}
The third one is RH parameterization with four parameters
\begin{eqnarray}
f^+(q^2) = \frac{f(0)(1-\delta \cdot \tilde{q}^2)}
  {(1-\tilde{q}^2)(1-\tilde{q}^2/\gamma)}. 
\end{eqnarray}
For all three cases the parameterization is the same for $f^0$,
First, the momentum dependent form factor data is interpolated to
fixed $E_{\pi}$'s  using these parameterization. Second, the chiral
limit is taken for each fixed  $E_{\pi}$ using staggered chiral
perturbation  theory~\cite{Aubin:2005aq}. It turns The results with 
different parameterizations are very well consistent with each other.
Choosing BZ parameterization for the best result, they obtain 
\begin{eqnarray}
\frac{1}{|V_{ub}|^2} \int_{16 GeV^2}^{q^2_max} \frac{d\Gamma}{d q^2}
= 1.46(23)(27) \mbox{ psec$^{-1}$}\nonumber
\end{eqnarray}
Using the experimental data from HFAG~\cite{ref:HFAG}
 $Br(q^2 > 16 GeV^2)=0.40(4)(4)\times 10^{-4}$, it leads to 
\begin{eqnarray}
|V_{ub}|=4.22(30)(51) \times 10^{-3},
\end{eqnarray}
which should be compared with FNAL/MILC results~\cite{Okamoto:2004xg},
\begin{eqnarray}
|V_{ub}| = 3.76(25)(65) \times 10^{-3}
\end{eqnarray}  
(See Fig. 8). 
Table~\ref{tab:Gam_16} shows the partial branching fraction for 
$q^2 > 16 GeV^2$ for various lattice calculations. So far within large 
errors, all the results are consistent. The average of $n_f=2+1$
results seems somewhat smaller than that of $n_f=0$ but not
significantly. 
\begin{table}
\begin{center}
\begin{tabular}{llll}
\hline
$n_f$ & Group& heavy  
& $\frac{\Gamma (q^2>16 GeV^2)}{|V_{ub}|^2}$ ps$^{-1}$\\
\hline
0     & UKQCD~\cite{Bowler:1999xn}
& clover   &  2.30($^{+77}_{-51}$)(51)      \\
0     & APE~\cite{Abada:2000ty}
     & clover   &  1.80($^{+89}_{-71}$)(47)      \\
0     & FNAL~\cite{El-Khadra:2001rv}
    & fermilab &  1.91($^{+46}_{-13}$)(31)    \\
0     & JLQCD~\cite{Aoki:2001rd}
   & NRQCD     &  1.71(66)(46)           \\
\hline
2+1   & HPQCD/MILC~\cite{Gulez:2006dt}  & NRQCD    & 1.46(23)(27)\\ 
2+1   & FNAL/MILC~\cite{Okamoto:2004xg} & fermilab & 1.83(50)   \\ 
\hline
\end{tabular}
\caption{ Values for partial branching fraction  $\frac{\Gamma
(q^2>16 GeV^2)}{|V_{ub}|^2}$  ps$^{-1}$ for various  lattice
calculations.  }  
\label{tab:Gam_16}
\end{center}
\end{table}

\begin{figure}[h]
\begin{center}
\includegraphics[width=8cm]{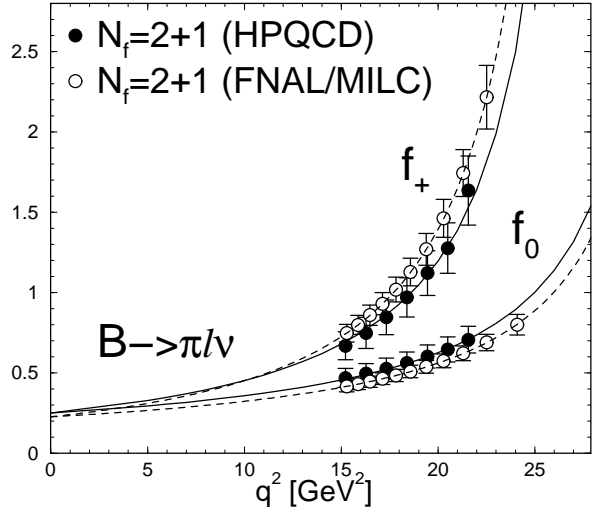}
\caption{$B\rightarrow \pi l\nu$ form factors in $n_f=2+1$ QCD.}
\end{center}
\label{fig:B2pi}
\end{figure}

\section{$m_b$}
Alpha collaboration made a quenched study of $1/M$ corrections to
HQET, which is an update of last years work.
Matching of QCD and HQET at small volume, step scaling in HQET towards
larger volume and computation of $m_{B_s}$ in large volume and finally 
convert to $m_b$. 
Last year they quoted that value
\begin{eqnarray}
m_b^{\overline{MS}}= m_b^{stat} + m_b^{(1)}, &\\
m_b^{stat}=4.350(64) \mbox{GeV}, &  
m_b^{(1)}  =0.049(29) \mbox{GeV} 
\end{eqnarray}

Guazzini et al.~\cite{Guazzini:2006bn}
computed the bottom quark mass using similar finite size scaling as $f_B$. 
Their preliminary results are 
\begin{eqnarray}
m_b^{RGI}= 6.96(11)& \mbox{ GeV} & \mbox{(only Rome II)}, \\
m_b^{RGI}= 6.89(11)& \mbox{ GeV} & \mbox{(Static + Rome II)} .
\end{eqnarray}

Kronfeld and Simone~\cite{Kronfeld:2000gk}
 made a quenched study of HQET parameter
$\bar{\Lambda}$, $\lambda_1$, and $m_b$. 
The idea is that HQET relation the heavy-light meson mass can be
expressed as
\begin{eqnarray}
M(m) = m + \bar{\Lambda} +\frac{\lambda_1}{2m} 
- d_J \frac{\lambda_2}{2m} + O(1/m^2).
\end{eqnarray}
Fitting the mass dependence of the heavy-light meson from lattice
calculation one can extract $\bar{\Lambda}$, $\lambda_1$, and
$m_b$ in lattice scheme. Using perturbation theory one can then
convert HQET parameters to another short distance scheme free from
renormalon ambiguities. Application of this method by to $n_f=2+1$
unquenched QCD by Fermilab
collaboration is reported in this conference~\cite{ref:Freeland}.

\begin{table}[h]
\begin{center}
\begin{tabular}{llll}
\hline
\hline
$n_f$ & Renormalization & Group & $\bar{m}_b(\bar{m}_b)$ (GeV)\\
\hline
\multicolumn{4}{l}{ $m_{B_s}$ and HQET} \\
\hline
 0    &  NNLO     & Martinelli, Sacrahjda~\cite{Martinelli:1998vt}
  & 4.38(5)(10)\\
      & Non pert. & Della Morte et al.~\cite{DellaMorte:2005gc}
     & 4.350(64) \\
\hline
 2    & NNNLO    & Renzo et al.~\cite{DiRenzo:2004xn}
            & 4.21(3)(5)(4)\\ 
      & NNLO     & McNeile et al.~\cite{McNeile:2004cb}
         & 4.25(2)(10)\\ 
\hline
\multicolumn{4}{l}{ $\Upsilon$ and NRQCD} \\
\hline
 2+1  & NLO    & Gray et al.~\cite{Gray:2005ur}
             & 4.4(3)\\ 
      & NLO    & Nobes, Trottier~\cite{Nobes:2005dz}
           & 4.7(4)\\
\hline
\hline
\end{tabular}
\caption{$\bar{m}_b(\bar{m}_b)$}
\end{center}
\end{table}
Recently HQET parameters are extracted by the global fit of the
various moments for inclusive B decays  such as 
$\langle E_l^n \rangle$, $\langle m_X^{2n} \rangle$ in  $B\rightarrow
X_c l\nu$ or $\langle E_\gamma^n \rangle$, in  $B\rightarrow X_s \gamma$ .
The result~\cite{Buchmuller:2005zv}
 is 
\begin{eqnarray}
m_b^{\overline{MS}}= 4.20(2)(5) \mbox{ GeV}\nonumber,
\end{eqnarray}

where the first and the second errors are the experimental and
theoretical errors. These determination is used to improve the
accuracy of $|V_{cb}|$ and $|V_{ub}|$ determinations from inclusive 
semileptonic decays. A better determination of HQET parameters would 
provide further improvement in  $|V_{cb}|$ and $|V_{ub}|$
determinations, which will be possible near future. Summary of recent
results are given in Table 9. 

\section{New methods}
\subsection{Dispersive bounds for form factors}

The momentum range of $B\rightarrow \pi l\nu$ form factors computed 
from Lattice QCD is limited by the small recoil or large $q^2$
region. This leads to a big disadvantage because most of the
experimental data lies in large recoil region. While one can
extrapolate in $q^2$ with a fit ansatz, this will always introduce 
some model dependence. Dispersive bounds is one possible way to
constrain the $q^2$ dependence in model independent fashion using
unitarity.

\begin{figure}[h]
\begin{center}
\includegraphics[width=8cm]{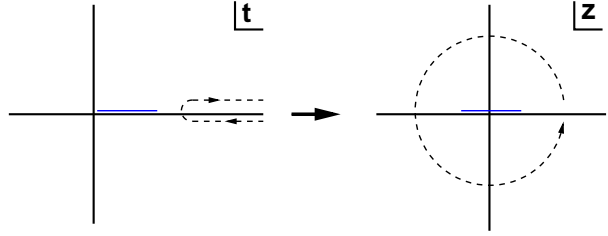}
\caption{A map from $t$ plane to $z$ plane}
\end{center}
\label{fig:map}
\end{figure}
Consider the imaginary part of the vacuum polarization amplitude for
the current $V(x)=\bar{u}\gamma_{\mu}b(x)$ and a map as in 
Fig. 9
\begin{eqnarray}\label{eq:ope}
\!\!\Pi^{\mu\nu}(q) 
& \equiv & i  \int d^4x\, e^{iq\cdot x} \langle 0 | T\left\{ V^\mu(x)
V^{\nu\dagger}(0) \right\} | 0 \rangle  \nonumber\\ 
&=& (q^\mu q^\nu - g^{\mu\nu} q^2)\Pi_1(q^2) + q^\mu q^\nu \Pi_0(q^2) \,, \\
z(t,t_0)& = &\frac{\sqrt{t_+-t}-\sqrt{t_+-t_0}}
          {\sqrt{t_+-t}+\sqrt{t_+-t_0}},  \ (\mbox{ with }
          t_{\pm}=(m_B\pm m_{\pi})^2 ),  
\end{eqnarray}

Then,  from dispersion relations one obtains 
\begin{eqnarray}
\chi_{F_+}(Q^2) &=& {1\over 2}{\partial^2 \over \partial (q^2)^2} \left[ q^2 \Pi_1 \right] 
= {1\over \pi} \int_0^\infty\! dt\, {t {\rm Im}\Pi_1(t) \over (t+Q^2)^3 } \,, \nl
\chi_{F_0}(Q^2) &=& {\partial \over\partial q^2} \left[ q^2 \Pi_0 \right] 
= {1\over \pi}  \int_0^\infty\! dt\, {t {\rm Im}\Pi_0(t) \over
(t+Q^2)^2 } \,. 
\label{eq:disp}
\end{eqnarray} 
with $Q^2= -q^2$ and $\eta$ an isospin factor, while $\chi$ 's can be
computed using OPE and perturbative QCD. 
Unitarity tells us that this is equal to the sum over all the hadronic
states. and dropping all   the excited states and leaving only the B
$\pi$ state gives an exact bound. 
\begin{eqnarray}
{\eta \over 48\pi}{ [(t-t_+)(t-t_-)]^{3/2} \over t^3} |F_+(t)|^2
&\le& {\rm Im} \Pi_1(t) \,, \nl
{\eta t_+ t_- \over 16\pi}{ [(t-t_+)(t-t_-)]^{1/2} \over t^3} |F_0(t)|^2 &\le& 
{\rm Im} \Pi_0(t) \,, 
\label{eq:bound}
\end{eqnarray} 
shows that an upper bound on the norm can be established 
by choosing [recall that $|z|=1$ along the integration contour in
(\ref{eq:disp})] 
\begin{eqnarray}
\phi_+(t,t_0) &=& \sqrt{\eta\over 48\pi}  
{t_+ - t \over (t_+ - t_0)^{1/4} } 
\left( z(t,0)\over -t \right) 
\left( z(t,-Q^2) \over -Q^2 - t \right)^{3/2}
\left( z(t,t_0) \over t_0 - t \right)^{-1/2} 
\left( z(t,t_-) \over t_- - t \right)^{-3/4} \,, 
\nl
\phi_0(t,t_0) &=& \sqrt{\eta t_+ t_- \over 16\pi} 
{\sqrt{t_+ - t} \over (t_+ - t_0)^{1/4} } 
\left( z(t,0) \over -t \right) 
\left( z(t,-Q^2) \over -Q^2 - t \right)
\left( z(t,t_0) \over t_0 - t \right)^{-1/2} 
\left( z(t,t_-) \over t_- - t \right)^{-1/4}  \,. 
\label{eq:phi}
\end{eqnarray} 

Combining Eqs.~\ref{eq:disp},~\ref{eq:bound}, ~\ref{eq:phi} 
and making change of variables in the integration from $t$ to $z$.
We obtain 
\begin{eqnarray}
\langle \phi f_0 | \phi f_0 \rangle < \chi_0, & 
\langle P \phi f_+ | P \phi f_+ \rangle < \chi_+,
\end{eqnarray}
where J is a quantity which can be obtained using OPE and perturbative
QCD. The inner product $\langle g | h \rangle$ for arbitrary functions
$g(z)$ and$h(z)$  is defined by the integral along the unit circle in
$z$ plane as  
\begin{eqnarray}
\langle g | h \rangle \equiv \int \frac{dz}{2\pi i} (g (z))^{\ast}.
\end{eqnarray}
$P(z)=z(t,m_B^*)$ is multiplied to $f_+$ in order  to remove $B^*$ pole
inside the unit circle.  Cauchy's theorem tells that if we know
additional integrated quantity $\langle g_i | P \phi_+ f_+ \rangle$ 
with a set of known functions $\{g_i(z), i=1,...,N \}$ one can make the
bound stronger as  
\begin{eqnarray}
\det \left( 
\begin{array}{cccc}
 \chi     &\langle P\phi_+ f_+ | g_1 \rangle & \ldots 
       &\langle P\phi_+ f_+ | g_N  \rangle \\
        \langle g_1  | P\phi f_+ \rangle 
       &\langle g_1 | g_1 \rangle & \ldots
       &\langle g_1 | g_N \rangle \\
\vdots &\vdots & \ddots & \vdots\\
        \langle g_N  | P\phi_+ f_+ \rangle 
       &\langle g_N | g_1 \rangle & \ldots
       &\langle g_N | g_N \rangle \\
 \end{array}
\right) > 0 .
\end{eqnarray}
Choosing $g_n(z) = \frac{1}{z-z(t)}$, Lellouch~\cite{Lellouch:1995yv}
 obtained stronger form factor
bounds with statistical analysis. Fukunaga and Onogi~\cite{Fukunaga:2004zz}
 improved the
bound using also the experimental $q^2$ spectrum as additional inputs. 
Arnesen et al. ~\cite{Arnesen:2005ez}
set $g_n(z) = z^n$ so that they can obtain the bound 
on the coefficients of the polynomial parameterization of the form
factor $\phi(z) f(z) = \sum_{n=0}^{\infty} a_n z^n$ as 
\begin{eqnarray}
\sum_{n=0}^{\infty} |a_n|^2 < \chi. 
\end{eqnarray}
This lead to a great simplification of the problem, although 
in practice one should truncate the polynomial at finite order 
so that the one has take into account this truncation error as 
the systematic error. 
Becher and Hill~\cite{Becher:2005bg},~\cite{Hill:2006ub}
 improved Arnesen et al's approach by imposing HQET
power counting to give stronger constraint than unitarity. Assuming
that this power counting argument correct they showed that only a few
degrees in polynomial is sufficient to approximate the form factor.
This statement is so far consistent with the observation 
from the Babar's data in Fig. 10. Of course one has
to bear in mind that with finite set of data one cannot always exclude
the possibility that the $q^2$ spectrum ( $z$ spectrum ) has yet
unobserved wiggly behavior from higher order terms in the polynomial
beyond our experimental resolution, but it will become more clear as 
experimental data will increase.

\begin{eqnarray}
\frac{1}{\chi}\sum_{n=0}^{\infty} |a_n|^2 < O((\Lambda/m_b)^3)
\mbox{  Becher-Hill's bound from HQET counting}
\end{eqnarray}

\vspace{0.5cm}
\begin{figure}[h]
\begin{center}
\includegraphics[width=8cm]{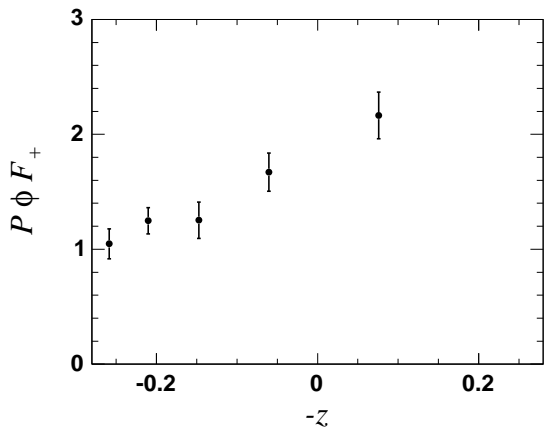}
\caption{Plot of form factor $f^+$ extracted from BaBar experimental
data multiplied by a function $P \phi$ as a function of $z$. It seems 
to be consistent with almost linear behavior in $z$. Figures from 
~\cite{Hill:2006ub}.}  
\end{center}
\label{fig:z dependence}
\end{figure}

Fermilab collaboration is carrying out an analysis based on Becher-Hill's
idea~\cite{ref:Ruth}. 

\subsection{all-to-all propagators for heavy-light meson}

TrinLat collaborations~
proposed to construct all-to-all propagators 
by combining low mode averaging~\cite{DeGrand:2002gm},~\cite{Giusti:2004yp}
 and random noise vector technique.
The noise should be diluted in time, spin and color sources.
They have shown that their all-to-all propagator is particularly useful 
for the heavy-light propagator with 20 eigen modes and single random source 
per dilution for each configuration. This method seems very promising. 
More experience in large volume is needed.

\section{Summary}
Experimental data are offering us a chance to overconstrain CKM.
Basic quantities such as decay constant, the bag parameter, form
factors , b quark masses are important in many ways.
Several different heavy quark formalism are useful for precision
calculation are studied.  New theoretical or calculation methods are
proposed to give better accuracy. 

\section*{Acknowledgments}
I would like to thank T.-W.~Chiu, C.N.H.~Chirst, S.~Hashimoto,
J.~Heitger, D.~Guazzini, A.S.~ Kronfeld, H.-W.~Lin, S.~Ohta,  S. Ryan,
J.~Shigemitsu, J.~N. Simone, R.~Sommer,  N. Tantalo, R. Van de Water,
N.~Yamada for providing me useful information and/or having
discussions for  my talk.  I also thank C. Davies,
C.~McNeile, T.~Matsuki, G.~von Hippel,  X.-Q.~Luo for informing me
about their work on spectrum and matrix elements in heavy quark
physics. I apologize that I could not cover those topic in my talk.    


\end{document}